\documentstyle[aps,12pt]{revtex}

\input psfig
\begin{document}

\centerline{August 19, 1994}
\bigskip

{\bf Spin wave frequency and critical dynamics of ferromagnets below
$T_{\rm c} \, ^{*)}$}
\bigskip

H.\ Schinz{$^{a)}$}, F.\ Schwabl{$^{a)}$}

$^{a)}$ Inst.\ f.\ Theor.\ Physik,
  TU M\"unchen, D-85747 Garching, Germany
\bigskip\bigskip

{\bf Abstract:}

Employing mode-coupling theory we show that the amplitude of the spin wave
frequency scaling function for the isotropic Heisenberg Hamiltonian is
universal. Theoretical and experimental values for Fe, Ni, Co, EuO, and EuS
agree quite well. Recent measurements of the longitudinal line width in Ni are
explained quantitatively.
\bigskip
\bigskip

\bigskip
\bigskip

{\bf Further correspondence to:}

{\obeylines
H.\ Schinz
Institut f\"ur Theoretische Physik
Technische Universit\"at M\"unchen
James-Franck-Stra\ss e
D-85747 Garching
Germany

e-mail: schinz@physik.tu-muenchen.de
FAX: 49-(0)89-$\,$3209-2296}
\bigskip
\bigskip

$^{*)}$ This work has been supported by the
German Federal Ministry for Research and Technology (BMFT) under contract
number 03-SC3TUM

\newpage
In contrast to the very satisfactory situation above $T_{\rm c}$, where
quantitative
agreement between experiment and theory in the description of critical
dynamics has been achieved
(see e.g.\ \cite{FreySchwabl_88}, \cite{aboveT_c} and ref.\ therein),
the situation in the ferromagnetic phase is far less clear.
Even in the isotropic case there are open questions. The
Hamilton operator we deal with reads
\begin{equation}
  H = \int\limits_{\vec q} S_{\vec q}^\alpha
    \left(J_0 + J q^2 \right) \delta^{\alpha \beta}
    S_{-{\vec q}}^\beta \qquad .
\end{equation}
A very successful
theory to describe critical dynamics in magnets has been mode-mode-coupling
(MC) theory \cite{FreySchwabl_88}.
Below $T_{\rm c}$ an important ingredient of this theory is the so called
frequency
matrix, which basically determines the eigenfrequencies of the fundamental
excitations -- the spin waves. To describe critical phenomena it is essential
to introduce scaling functions for the various entities that appear in the
theory. If we use the scaling variable $x = 1/q\xi$ where $q$ is the wave
vector and $\xi$ the correlation length, temperature dependent according to
$\xi = \xi_0 \tau^{-\nu}$, $\tau = 1 - T/T_{\rm c}$, the spin wave
frequency can be shown \cite{FreySchwabl_88} to have the form
\begin{equation}
  \omega_{\vec q}(\xi) = F q^z \cdot \hat\omega(x) \quad , \qquad
  \hat\omega(x) = \hat b \sqrt{x} \qquad ,
  \label{omeg_skal}
\end{equation}
valid for all $x$.
Here, $z$ is the (universal) dynamic critical exponent
and $F$ is a nonuniversal amplitude, setting the scale for dynamics.
The question now arises whether the scaling function $\hat\omega$ is
universal, i.e.\
whether the amplitude $\hat b$ is universal. Furthermore its value is
of importance. By a general argument we
will show that based on MC theory and the dynamical scaling hypothesis
$\hat\omega$ should in fact be universal. Analysing the scaling properties
of MC theory we derive an expression relating $\hat b$ to other universal
amplitude ratios. This is used to determine theoretical values for
$\hat b$, which will then be compared with experimentally derived values
from Fe, Ni, Co, EuO, and EuS. Finally, we discuss briefly the consequences
for the results of MC theory for the line width scaling functions.

The dynamical scaling hypothesis \cite{SS_HH} for the dynamic susceptibility
can be stated in the following form \cite{HH_review}
\begin{equation}
  \chi(\vec q, \omega) = \chi_{\vec q} \cdot
    Y({\scriptstyle{\omega\over F q^z},{1\over q\xi}}) \quad , \qquad
  \omega_\Psi(\vec q) = F q^z \cdot
    \bar\omega({\scriptstyle{1\over q\xi}}) \qquad .
  \label{chi_skal}
\end{equation}
$\chi_{\vec q}$ is the static susceptibility, $\omega_\Psi$ a characteristic
frequency, $Y$ and $\bar\omega$ are
universal scaling functions. In MC
theory a Kubo relaxation function $\Phi$ is introduced which is related to
$\chi$ via $
  \Phi(\vec q,\omega) =
    \left\{ \chi(\vec q, \omega) - \chi(\vec q, 0) \right\}
    / i \omega$.
If now we define a new function $\Gamma$ related to $\Phi$ by
\begin{equation}
  \Phi(\vec q,\omega) = i \sqrt{\chi_{\vec q}}
    {1\over \omega 1 + \omega_{\vec q} + i \Gamma(\vec q, \omega)}
    \sqrt{\chi_{\vec q}}
  \label{gam_skal}
\end{equation}
then, obviously, the scaling law (\ref{chi_skal}) for $\chi$ implies a
corresponding scaling
law for $\Phi$ and for $\Gamma - i \omega_{\vec q}$ with a universal scaling
function $G$
for this last quantity. To show that both functions $\omega_{\vec q}$ and
$\Gamma$
are universal seperately we use MC theory which expresses the linewidth
$\Gamma$ in terms of $\Phi$ through
$
  \Gamma(\vec q, t) =
    4k_BT \int_{\vec p} v(\vec q, \vec p)
    \Phi(\vec p, t) \Phi(\vec q - \vec p, t)
$
with some kernel \hbox{${v({\vec q}, {\vec p})}$} \cite{FreySchwabl_88}.
If we employ the form (\ref{omeg_skal}) for $\omega_{\vec q}$
with a nonuniversal $\hat b$, and solve for $\Gamma$ (see below), then $G$
still depends on the (assumed nonuniversal) $\hat b$
in contradiction to dynamical scaling. We therefore conclude that $\hat b$
is universal. This implies also that the scaling function for $\Gamma$ is
universal.

Analysing the scaling properties
of the MC equations and combining this with the static and dynamic scaling
hypothesis \cite{Privman_91} we find the final result for $\hat b$
\cite{Schinz}
\begin{equation}
  \hat b \; =  \;
    {2\pi^2 \over 5.13} \left( {T_{\rm c}\over T} \right)^{1\over 2}
    \left( {R_{\rm c}\over (R_\xi^+)^d} \right)^{1\over 2}
    \left( {\xi_0\over\xi_0^T} \right)^{d-2}
    \left( {\xi_-\over\xi_+} \right)^{z-2} \qquad ,
  \label{hat_b}
\end{equation}
At $T_c$ the prefactor becomes a universal numerical constant,
$R_{\rm c}$ and
$R_\xi^+$ are universal ratios of nonuniversal amplitudes as defined in Ref.\
\cite{Privman_91}, $\xi_0^T$ is a transversal correlation length below
$T_{\rm c}$
\cite{Privman_91}, and $\xi_+$, $\xi_-$ are (longitudinal) correlation lengths
above and below $T_{\rm c}$ respectively.
Again we conclude that $\hat b$ is universal and determined by other universal
ratios of nonuniversal amplitudes. We also
see that this amplitude depends only on static quantities in correspondence
to the dynamic exponent $z$, which for model J is given by static exponents
($z = d/2 + 1 - \eta/2$).

In Ref.\ \cite{Schwabl_71} a Green's function formalism
and sum rules were employed to derive the scaling form for $\omega_{\vec q}$
together with the prefactor
\begin{equation}
  \hat b \approx \pi^{3\over 2}/5.13 \approx 1.08 \qquad .
\end{equation}
This value has also been used in consecutive applications of MC theory on
magnets (\cite{FreySchwabl_88} and ref.\ therein). A lot of work has been done
on the calculation of several
universal amplitude ratios for the isotropic n-vector model (cf.\
\cite{Privman_91}). Using these to evaluate $\hat b$ from Eq.\ (\ref{hat_b}) we
get
\begin{equation}
  \hat b = 1.9 \pm 0.4 \qquad .
\end{equation}
In principle, the values for all these universal quantities depend on the fixed
point in the renormalisation group sense, i.e.\ could change by going from
the isotropic to the dipolar fixed point. But the findings for other quantities
\cite{Privman_91} suggest that the numerical values will not be too different.

For the experimental determination of $\hat b$ one plots the characteristic
frequency $\omega_{\rm ch}$ normalised by the damping at $T_{\rm c}$ vs.\
scaling variable
$x$. For large $x$ (small $q$) we get $\omega_{\rm ch} \approx
\omega_{\vec q}$ and therefore
\begin{equation}
  {\omega_{\rm ch}(\vec q,\tau)\over\Gamma_{\vec q}
    (\tau\raise.3ex\hbox{$\scriptscriptstyle\rightarrow$}0)}
    \approx W_- \; \sqrt{1\over q\xi_-} \approx {\hat\omega(x)\over G(0)}
    \quad , \qquad
  \hat b = G(0) \; W_- \qquad ,
\end{equation}
where $G(0)$ accounts for the normalisation of the scaling functions.
The resulting values of $\hat b$ for the different substances are given in
table \ref{table}. Depending on which experiment \cite{experimente}
one uses one gets slightly different values of $\hat b$.
\bigskip

\begin{table}[h]
\caption{Experimental values for the spin wave amplitude $\hat b$ from
Ref.\ \protect\cite{experimente}.}
\bigskip\bigskip
\begin{tabular}{lccccc}
 &$Fe$ &$Ni$ &$Co$ &$EuO$ &$EuS$\\
 \hline
 $\hat b \qquad$ &1.5(1) &1.5(1) &1.6(2) &1.3(2) &1.4(2)\\
                 &1.8(1) &2.1(1) &       &       &1.9(3)\\
\end{tabular}
\label{table}
\end{table}

If we solve the MC equations within the Lorentzian approximation (cf.\
\cite{FreySchwabl_88}) we get $\hat b$-dependent results for the scaling
function of the
longitudinal linewidth $g_{\rm L}$ (\hbox{Fig.\ \ref{fig}}). Included are also
the recently
obtained experimental results for Ni \cite{Boeni_91}.
Obviously, the agreement can be improved considerably
if we choose the appropriate value of $\hat b$ \cite{zobel_91}.
Instead of the monotonic increase found for the RPA value 1.08
\cite{FreySchwabl_88} we now get a minimum in the scaling function.
Thus, we are able to explain the data for the longitudinal scaling function of
Ni quantitatively.
\bigskip

\begin{figure}[h]
  \psfig{figure=gl_fig.ps,width=12cm}
\caption{Scaling function $g_{\rm L}$ vs.\ $x = 1/q\xi$.
Experimental data from Ref.\ \protect\cite{Boeni_91}.}
  \label{fig}
\end{figure}

The authors would like to thank P.\ B\"oni for providing them with the
numerical values of the measurements from Ref.\ \cite{Boeni_91} and for some
valuable discussions. This work has been supported by the
German Federal Ministry for Research and Technology (BMFT) under contract
number 03-SC3TUM.

\end{document}